# On the Design and Invariants of a Ruled Surface

**Ferhat Taş**
Ph.D., Research Assistant.
Department of Mathematics, Faculty of Science, Istanbul University, Vezneciler, 34134, Istanbul,Turkey.
tasf@istanbul.edu.tr

**ABSTRACT**

*This paper deals with a kind of design of a ruled surface. It combines concepts from the fields of computer aided geometric design and kinematics. A dual unit spherical Bézier-like curve on the dual unit sphere (DUS) is obtained with respect the control points by a new method. So, with the aid of Study [1] transference principle, a dual unit spherical Bézier-like curve corresponds to a ruled surface. Furthermore, closed ruled surfaces are determined via control points and integral invariants of these surfaces are investigated. The results are illustrated by examples.*

**Classification:** 53A17-53A25

## 1 Introduction

Many scientists like mechanical engineers, computer scientists and mathematicians are interested in ruled surfaces because the surfaces are widely using in mechanics, robotics and some other industrial areas. Some of them have investigated its differential geometric properties and the others, its representation mathematically for the design of these surfaces. Study [1] used dual numbers and dual vectors in his research on the geometry of lines and kinematics in the 3-dimensional space. He showed that there exists a one-to-one correspondence between the position vectors of DUS and the directed lines of space $\mathbb{R}^3$. So, a one-parameter motion of a point on DUS corresponds to a ruled surface in 3-dimensional real space.

Hoschek [2] found integral invariants for characterizing the closed ruled surfaces. Furthermore, Gürsoy [3] studied some relationship finding between the dual integral invariant and the real integral invariants of a closed ruled surface. He concerned the motion of a point on DUS which draws a closed trajectory. However, these studies still remain theoretical, except for some examples.

The purpose of investigation of ruled surfaces is to find its computational properties since it is used in design and manufacturing. So, we have found a useful method to the representation of a ruled surfaces to designing it. For this, we need a designable curve on DUS.

Spherical Bézier curves were introduced by Shoemake [4] using the spherical interpolation which is called by Slerp. This interpolation method is valid for two quaternions. He developed a method to get a spherical Bézier curve via combining the spherical curve segments continuously. P. Crouch, G. Kun, F. Silva Leite [5] have studied on this subject. Q.J. Ge and B. Ravani [6, 7] developed a computational geometric structure by using dual quaternion interpolation for continuous Bézier motion of the rigid body. Y. Zhou, J. Schulze and S. Schäffler [8] studied about optimization on DUS and defined the dual spherical spline for the design of a blade surface as ruled surface to cutting and shaping the shoes. In [9], In this study, the dual unit spherical Bezier curve was created by the global interpolation method of equally spaced points on the dual sphere. Of course, this has led to some restrictions. Thus, this curve thereby forming a negative in terms of ability to flexibly design the ruled surface.

It is well-known that dual curves on DUS have an important role in kinematics, design and control problems. In the theory of mechanisms, Bézier curves and surfaces provide very useful results. The main purpose of Bézier technic is to present the algorithmic basis to applicable computer design. So, the aim of this study is to generate any differentiable dual unit spherical Bézier curve and investigate the properties of corresponding ruled surfaces in terms of control points.

Moreover, our method represents a flexible design since control points provide flexibility in practice i.e. mechanisms. For this aim, we use a map from the domain of the real unit sphere (RUS) to its surface to define a dual unit spherical Bézier-like curve. In this case, a dual unit spherical Bézier-like curve can be obtained by using the control points on the RUS.

**2 Basic Concepts**

This section is about fundamentals of the kinematics that generate a ruled surface and its representation via dual unit vectors.

**2.1 Dual Vector Representation of a Ruled Surface**

Clifford [10] defined dual numbers as a tool for his geometrical studies. The set of dual numbers is $\mathbb{D} = \{A = a + \varepsilon \bar{a} | a, \bar{a} \in \mathbb{R}, \varepsilon^2 = 0\}$.

Here $\varepsilon$, namely, the dual unit can be represented by an ordered pair $\varepsilon$ = (0, 1). This set forms an associative ring over real numbers with the following operations,

$$A + B = a + \varepsilon \bar{a} + b + \varepsilon \bar{b} = a + b + \varepsilon(\bar{a} + \bar{b})$$

and

$$AB = (a + \varepsilon\bar{a})(b + \varepsilon\bar{b}) = ab + \varepsilon(\bar{a}b + a\bar{b}).$$

Let $f$ be a differentiable function with respect to the dual variable $X = x + \varepsilon\bar{x}$. Then, by using the Taylor series of $f$, we can write the following equation [1],

$$f(X) = f(x + \varepsilon\bar{x}) = f(x) + \varepsilon\bar{x}\frac{df(x)}{dx}.$$

We can give some examples:

$\sin(x + \varepsilon y) = \sin(x) + \varepsilon y\cos(x)$,

$\cos(x + \varepsilon y) = \cos(x) - \varepsilon y\sin(x)$,

$\exp(x + \varepsilon y) = \exp(x) + \varepsilon y\exp(x)$.

After Clifford's works, Study used the dual numbers and the dual vectors in his research on the geometry of lines and kinematics. The set of dual vectors is defined as

$$\mathbb{D}^3 = \{\mathbf{X} = \mathbf{x} + \varepsilon\bar{\mathbf{x}} | \mathbf{x}, \bar{\mathbf{x}} \in \mathbb{R}^3, \varepsilon^2 = 0\}.$$

In $\mathbb{D}^3$, a $\mathbb{D}$ – valued bilinear form is defined by

$$\langle \mathbf{U}, \mathbf{V} \rangle = \langle \mathbf{u}, \mathbf{v} \rangle + \varepsilon(\langle \mathbf{u}, \bar{\mathbf{v}} \rangle + \langle \bar{\mathbf{u}}, \mathbf{v} \rangle)$$

where $\mathbf{U} = \mathbf{u} + \varepsilon\bar{\mathbf{u}}$ and $\mathbf{V} = \mathbf{v} + \varepsilon\bar{\mathbf{v}}$. Then we can define a dual unit vector as follow

$$\|\mathbf{U}\|^2 = \langle \mathbf{U}, \mathbf{U} \rangle = \|\mathbf{u}\|^2 + 2\varepsilon\langle \mathbf{u}, \bar{\mathbf{u}} \rangle = 1.$$

So, $\mathbf{U}$ is a unit vector iff $\|\mathbf{u}\| = 1, \langle \mathbf{u}, \bar{\mathbf{u}} \rangle = 0$.

We can explain what the geometric meaning of scalar product of two dual vector $\mathbf{U}$ and $\mathbf{V}$ is: Let *U* and *V* two lines in 3-dimensional real space. Then the real and dual parts of the scalar product which is defined above measure the angle and the shortest distance between two lines, respectively. The direction vector of the common perpendicular line is found by **w** = **u**×**v**.

**Definition 1** The dual angle between the dual vectors **U** and **V** can be represented by

$\widehat{\sphericalangle}(\mathbf{U},\mathbf{V}) = \sphericalangle(\mathbf{u},\mathbf{v}) + \varepsilon \mathrm{d}(U,V)$.

where the dual angle between two lines, the angle between direction vectors and the shortest distance between two lines are denoted by $\widehat{\sphericalangle}(\mathbf{U},\mathbf{V})$, $\sphericalangle(\mathbf{u},\mathbf{v})$ and $\mathrm{d}(U,V)$, respectively.

Let $\mathbf{X}(t) = \mathbf{x}(t) + \varepsilon\bar{\mathbf{x}}(t)$ be a one-parameter dual unit curve. Then its correspondence in $\mathbb{R}^3$ w.r.t. Study's transference principle is given by

$\mathbf{r}(t,u) = \mathbf{c}(t) + u\mathbf{x}(t)$

where $t, u \in \mathbb{R}$. $\mathbf{c} = \mathbf{x} \times \bar{\mathbf{x}}$ and **x** are directrix curve and generator vector of the ruled surface, respectively.

## 2.2 Kinematic Generation of a Ruled Surface

In this section, following Blaschke [11] we give some differential geometric properties of a ruled surface which is drawn by a one-parameter motion in 3-dimensional real space.

Let $L(t)$ a one-parameter family of lines in 3-dimesional real space and defined by:

**c, x**: $D \subset E \to E^3$, **r**: $D \times \mathbb{R} \to E^3$,

$$\mathbf{r}(t,w) = \mathbf{c}(t) + w\mathbf{x}(t) \qquad (1)$$

where $\mathbf{c}(t)$ ve $\mathbf{x}(t)$ are base and generator curve, respectively. In addition, we assume $\|\mathbf{x}\| = 1$. Then the mapping $\mathbf{x}(D)$ be the spherical image of $L$. Since vector **c** is the position vector and **x** is the direction vector at a point $t_0$ on the line $L$, then the dual vector representation of a line $L$ is

$$\mathbf{X}(t) = \mathbf{x}(t) + \varepsilon \mathbf{c}(t) \times \mathbf{x}(\mathrm{t}) = \mathbf{x}(t) + \varepsilon\bar{\mathbf{x}}(t). \qquad (2)$$

as shown Fig. 1.

Let

$$\left\{ \mathbf{X}_1 = \mathbf{X}, \ \mathbf{X}_2 = \frac{\mathbf{X'}_1}{\|\mathbf{X'}_1\|}, \ \mathbf{X}_3 = \mathbf{X}_1 \times \mathbf{X}_2 \right\} \tag{3}$$

be a dual orthonormal frame. While $\mathbf{X}_1$ generate the ruled surface, we get the derivative equations of this motion like this;

$$\begin{bmatrix} \mathbf{X'}_1 \\ \mathbf{X'}_2 \\ \mathbf{X'}_3 \end{bmatrix} = \begin{bmatrix} 0 & \hat{\kappa} & 0 \\ -\hat{\kappa} & 0 & \hat{\tau} \\ 0 & -\hat{\tau} & 0 \end{bmatrix} \begin{bmatrix} \mathbf{X}_1 \\ \mathbf{X}_2 \\ \mathbf{X}_3 \end{bmatrix}$$

$$\hat{\kappa} = \hat{\kappa}(t) = \|\mathbf{X'}_1\|, \qquad \hat{\tau} = \hat{\tau}(t) = \frac{[\mathbf{X}_1, \mathbf{X}_1', \mathbf{X}_1'']}{\|\mathbf{X'}_1\|^2}$$

where $\hat{\kappa}$ and $\hat{\tau}$ are differentiable dual invariants and called by dual curvature and dual torsion of $\mathbf{X}_1(t)$-ruled surface, respectively. Here, if we expand this functions to the real and dual parts, we get

$$\hat{\kappa} = \kappa + \varepsilon \bar{\kappa}, \quad \hat{\tau} = \tau + \varepsilon \bar{\tau},$$

and then derivative equations become

$$\begin{bmatrix} \mathbf{x}_1' \\ \mathbf{x}_2' \\ \mathbf{x}_3' \end{bmatrix} = \begin{bmatrix} 0 & \kappa & 0 \\ -\kappa & 0 & \tau \\ 0 & -\tau & 0 \end{bmatrix} \begin{bmatrix} \mathbf{x}_1 \\ \mathbf{x}_2 \\ \mathbf{x}_3 \end{bmatrix} \tag{4}$$

$$\bar{\mathbf{x}}_1' = \bar{\kappa}\mathbf{x}_2 + \kappa\bar{\mathbf{x}}_2, \ \bar{\mathbf{x}}_2' = -\bar{\kappa}\mathbf{x}_1 + \bar{\tau}\mathbf{x}_3 - \kappa\bar{\mathbf{x}}_1 + \tau\bar{\mathbf{x}}_2, \ \bar{\mathbf{x}}_3' = -\bar{\tau}\mathbf{x}_2 - \tau\bar{\mathbf{x}}_2. \tag{5}$$

Since

$$\int \|\mathbf{X'}_1\| dt = \int \hat{\kappa} dt = \int \kappa dt + \varepsilon \int \bar{\kappa} dt$$

dual arc-length of $\mathbf{X}_1(t)$ dual spherical curve is an invariant of the ruled surface then $\int \kappa dt$ and $\int \bar{\kappa} dt$ are integral invariants of this surface. Analogously,

$$\int \|\mathbf{X'}_3\| dt = \int \hat{\tau} dt = \int \tau dt + \varepsilon \int \bar{\tau} dt$$

is dual arc-length of $\mathbf{X}_3(t)$ dual spherical curve thus, $\int \tau dt$ and $\int \bar{\tau} dt$ are integral invariants of this surface. So, we get real and dual parts of these functions like that:

$$\kappa = \|\mathbf{x'}_1\|,$$

$$\tau = \frac{[\mathbf{x}_1, \mathbf{x}_1', \mathbf{x}_1'']}{\|\mathbf{x'}_1\|^2}$$

$$\bar{\kappa} = \frac{\langle \mathbf{x'}_1, \bar{\mathbf{x}}'_1 \rangle}{\kappa}$$

$$\bar{\tau} = \frac{[\bar{\mathbf{x}}_1, \mathbf{x}_1', \mathbf{x}_1''] + [\mathbf{x}_1, \bar{\mathbf{x}}_1', \mathbf{x}_1''] + [\mathbf{x}_1, \mathbf{x}_1', \bar{\mathbf{x}}_1'']}{\kappa^2} - \frac{2\tau\bar{\kappa}}{\kappa}. \tag{6}$$

The distribution parameter of $\mathbf{X}_1(t)$ − ruled surface is given by

$$\delta = \frac{1}{d} = \frac{\langle \mathbf{x}_1', \bar{\mathbf{x}}_1' \rangle}{\mathbf{x}_1'^2} = \frac{\bar{\kappa}}{\kappa}, (\kappa \neq 0).$$

If $\kappa = 0$ then $\mathbf{X}_1(t)$ denotes a cylindrical surface.

We may give a base curve of $\mathbf{X}_1(t)$ − ruled surface like that,

$$\mathbf{c}(t) = \mathbf{x_1}(t) \times \bar{\mathbf{x}}_\mathbf{1}(t).$$

Then $\mathbf{X}_1(t)$ − ruled surface can be given in vector form:

$$\mathbf{r}(t, w) = \mathbf{x}_1(t) \times \bar{\mathbf{x}}_1(t) + w\mathbf{x}_1(t).$$

Let striction curve be $\mathbf{m}(t)$. Since $\mathbf{m}(t)$ is on the ruled surface and $\mathbf{x}_2$ is surface normal along striction curve then we can write the tangent vector of the striction curve of $\mathbf{r}$ ruled surface like

$$\frac{d\mathbf{m}}{dt} = a\mathbf{x}_1 + b\mathbf{x}_3, \quad a, b \in \mathbb{R}.$$

Besides, $\mathbf{m}(t)$ is the intersection point of $\mathbf{x}_1, \mathbf{x}_2, \mathbf{x}_3$ then,

$$\mathbf{m} \times \mathbf{x}_1 = \bar{\mathbf{x}}_1, \quad \mathbf{m} \times \mathbf{x}_3 = \bar{\mathbf{x}}_3$$

Differentiating these equations and with the help of the equation (5), we find

$$\frac{d\mathbf{m}}{dt} = \bar{\tau}\mathbf{x}_1 + \bar{\kappa}\mathbf{x}_3. \tag{7}$$

Then, the arc-length function $s(t)$ of the striction curve is found by

$$s(t) = \int \sqrt{\bar{\tau}^2 + \bar{\kappa}^2}\, dt.$$

The equation of the striction curve is given as follow

$$\mathbf{m}(t) = \mathbf{c}(t) - \frac{\langle \mathbf{x}_1'(t), \mathbf{c}'(t)\rangle}{\|\mathbf{x}_1'(t)\|^2}\mathbf{x}_1(t).$$

**2.3 Striction of a Ruled Surface**

From the fundamental theorem of space curves, we can determine a likely shape of a space curve from its curvature and torsion. If we want to determine the shape of a ruled surface (non-cylindrical) we have to know its Kruppa invariants; curvature, torsion and striction.

Let $s$ be arc-length of striction curve then,

$$\frac{d\mathbf{m}}{ds} = \cos\sigma(s)\mathbf{x}_1(s) + \sin\sigma(s)\mathbf{x}_3(s), \qquad (0 \leq \sigma(s) \leq 2\pi) \tag{8}$$

can be written where $\sigma(s)$ is striction of the ruled surface.

**Theorem 1** (G. Sannia, 1925)

Let $\kappa$ and $\sigma$ be continuously differentiable functions and $\tau$ be continuous on an interval $I \subset \mathbb{R}$. Then, these functions determine a twice continuously differentiable ruled surface, [12].

**Theorem 2** Let $\hat{\kappa}(t), \hat{\tau}(t)$ are $\mathbb{D}$-valued differentiable functions. Then these functions determine a dual curve $\mathbf{X}_1: I \to \mathbb{D}^3$ on an interval $I \subset \mathbb{R}$, [13].

**Corollary 1** Considering Study's transference principle, this theorem corresponds to that if we know $\mathbb{D}$-valued differentiable functions $\hat{\kappa}(t), \hat{\tau}(t)$ then we get only one ruled surface in 3-dimensional Euclidean space.

**2.4 Differential Geometric Properties of a Ruled Surface**

The Differential geometric properties of a ruled surface can be expressed in terms of $\kappa, \tau, \bar{\kappa}, \bar{\tau}$ functions.

Let the ruled surface be parametrized by (1). Then partial derivatives of the ruled surface

$\mathbf{r}_t = \bar{\tau}\mathbf{x}_1 + w\kappa\mathbf{x}_2 + \bar{\kappa}\mathbf{x}_3,$

$\mathbf{r}_w = \mathbf{x}_1, \quad \mathbf{r}_{ww} = 0,$  (9)

$\mathbf{r}_{tw} = \kappa\mathbf{x}_2$

may determine the unit normal of the ruled surface as follow

$$\mathbf{n} = \frac{\mathbf{r}_t \times \mathbf{r}_w}{\|\mathbf{r}_t \times \mathbf{r}_w\|} = \frac{(\bar{\tau}\mathbf{x}_1 + w\kappa\mathbf{x}_2 + \bar{\kappa}\mathbf{x}_3) \times (\mathbf{x}_1)}{\|(\bar{\tau}\mathbf{x}_1 + w\kappa\mathbf{x}_2 + \bar{\kappa}\mathbf{x}_3) \times (\mathbf{x}_1)\|} = \frac{\bar{\kappa}\mathbf{x}_2 - w\kappa\mathbf{x}_3}{\sqrt{(w\kappa)^2 + \bar{\kappa}^2}}.$$

Furthermore, we can achieve coefficients of metric.

$g_{11} = \langle \mathbf{r}_t, \mathbf{r}_t \rangle = \bar{\tau}^2 + (w\kappa)^2 + \bar{\kappa}^2$

$g_{12} = \langle \mathbf{r}_t, \mathbf{r}_w \rangle = \bar{\tau}$

$g_{22} = \langle \mathbf{r}_w, \mathbf{r}_w \rangle = 1$

are coefficients of the first fundamental form [I] and

$h_{11} = \langle \mathbf{r}_{tt}, \mathbf{n} \rangle = \dfrac{\bar{\kappa}(\bar{\tau} + w\kappa' - \tau) - w\kappa(w\kappa\tau + \bar{\kappa}')}{\sqrt{(w\kappa)^2 + \bar{\kappa}^2}}$

$h_{12} = \langle \mathbf{r}_{tw}, \mathbf{n} \rangle = \dfrac{\bar{\kappa}\kappa}{\sqrt{(w\kappa)^2 + \bar{\kappa}^2}}$

$h_{22} = \langle \mathbf{r}_{ww}, \mathbf{n} \rangle = 0$

are coefficients of the second fundamental form [II]. After some calculations, we get Gauss and mean curvatures of the ruled surface:

$$K_G(t,w) = \frac{h_{11}h_{22} - h_{12}^2}{g_{11}g_{22} - g_{12}^2} = -\frac{\bar{\kappa}^2 \kappa^2}{((w\kappa)^2 + \bar{\kappa}^2)^2},$$

$$K_M(t,w) = \frac{h_{11}g_{22} + g_{11}h_{22} - 2h_{12}g_{12}}{2(g_{11}g_{22} - g_{12}^2)}$$

$$= \frac{\bar{\kappa}(\bar{\tau} + w\kappa' - \tau) - w\kappa(w\kappa\tau + \bar{\kappa}') - 2\kappa\bar{\kappa}\bar{\tau}}{2((w\kappa)^2 + \bar{\kappa}^2)^{3/2}}.$$

**2.5 Integral Invariants of a Closed Ruled Surface**

In addition to definition of a one-parameter motion $\mathbf{X}_1(t)$, if it is periodic, i.e. $\mathbf{X}_1(t+P) = \mathbf{X}_1(t)$, then this motion to be closed on DUS. So, corresponding ruled surface is also closed. Hoschek defined integral invariants to this type surfaces: First integral invariant is pitch which is described like that: If $\mathbf{x}_1(t)$ generate a ruled surface and the motion is closed then, after one period, this vector is on the same line and generally different point. Length between these two points is called pitch of the ruled surface and it is formulated by

$$l_{\mathbf{X}_1} = -\oint \langle \mathbf{c}', \mathbf{x} \rangle dt. \tag{10}$$

Second integral invariant is angle of pitch which is described like that: Let $\boldsymbol{\xi}$ be a unit vector on the plane $(\mathbf{x}_2, \mathbf{x}_3)$ and generate a developable surface during this motion:

$$\boldsymbol{\xi} = \cos(\phi)\mathbf{x}_2 + \sin(\phi)\mathbf{x}_3$$

Then, the measure of total variation of the angle $\phi$ during this motion is called by angle of pitch of the ruled surface and formulated as follow:

$$\lambda_{\mathbf{X}_1} = \oint d\phi = \oint \langle \mathbf{x}_2', \mathbf{x}_3 \rangle dt. \tag{11}$$

**3. Construction of a Bézier-Like Curve on DUS**

DUS can be defined, similar to the real one, by

$$S_{\mathbb{D}}^2 = \{\mathbf{X} = \mathbf{x} + \varepsilon \bar{\mathbf{x}} | \|\mathbf{X}\| = 1, \mathbf{x}, \bar{\mathbf{x}} \in \mathbb{R}^3\}$$

Theoretically, a dual parametric representation of a DUS can be given in vector form [12]:

$$\mathbf{X}(\hat{u}, \hat{v}) = (\cos \hat{u} \sin \hat{v}, \sin \hat{u} \sin \hat{v}, \cos \hat{v}), \tag{12}$$

where $\hat{u} = u + \varepsilon \bar{u}, \hat{v} = v + \varepsilon \bar{v}, \ u, v, \bar{u}, \bar{v} \in \mathbb{R}$. Using Taylor expansion of dual-parameter functions, equation (12) becomes

$$\mathbf{X}(\hat{u}, \hat{v}) = \mathbf{x}(u, v) + \varepsilon[\bar{u}.\mathbf{x}_u(u, v) + \bar{v}.\mathbf{x}_v(u, v)]$$

or equivalently

$$\mathbf{X}(\hat{u}, \hat{v}) = (\cos u \sin v, \sin u \sin v, \cos v) + \varepsilon[\bar{u}(-\sin u \sin v, \cos u \sin v, 0) +$$

$$\bar{v}(\cos u \cos v, \sin u \cos v, -\sin v)]. \tag{13}$$

Equation (13) represents four-parameter subspace in real space and this is so-called a family of line complexes. The real part of the equation (13) depends on the $u, v$ parameters. Therefore, for the convenience of all parameters in the general equation (13), we take $\bar{u}, \bar{v}: \mathbb{R}^2 \mapsto \mathbb{R}, \bar{u} = \bar{u}(u, v)$ and $\bar{v} = \bar{v}(u, v)$. Then, the number of parameters descends from four to two. Thus, $\mathbf{X}$ indicates the two-parameter motion on DUS and so, we know this motion corresponds to a line-congruence in $\mathbb{R}^3$. Now, if we return to the main purpose of this paper, we will construct a one-parameter devisable motion on DUS via Bézier technic:

Let $\mathcal{B} = \{(u,v) \in \mathbb{R}^2 | \ 0 \leq u \leq \pi, \ 0 \leq v \leq 2\pi\}$ be a domain of RUS and $\{\mathbf{p}_0, \mathbf{p}_1, \ldots, \mathbf{p}_n\}$ is the set of the control points of degree n+1 Bernstein-Bézier curve on this domain where $\mathbf{p}_i \in (u,v)$. Bernstein-Bézier curve is then given by

$$\mathbf{b}(t) = \sum_{i=0}^{n} \binom{n}{i} (1-t)^{n-i} t^i \mathbf{p}_i, \qquad t \in \mathbb{R}. \tag{14}$$

It is known that if $\mathbf{p}_0 = \mathbf{p}_n$ and $\{\mathbf{p}_{n-1}, \mathbf{p}_0, \mathbf{p}_1\}$ are collinear then the curve has closed shape and $C^1$ continuity, respectively [13, 14]. Using these conditions, a closed Bézier curve $\mathbf{b}(t)$ can be obtained.

Mapping $\mathbf{b}(t)$ via $\mathbf{X}$, we get a closed spherical Bézier-like curve on RUS:

$$u = u(t) = \sum_{i=0}^{n} \binom{n}{i}(1-t)^{n-i} t^i \mathbf{p}_{i_1}, \ v = v(t) = \sum_{i=0}^{n} \binom{n}{i}(1-t)^{n-i} t^i \mathbf{p}_{i_2} \tag{15}$$

where $\mathbf{p}_{i_j}$ (j=1,2) are the jth coordinate of point $\mathbf{p}_i$. Since we already know functions $\bar{u}$ and $\bar{v}$, we get a closed dual spherical Bézier-like curve on DUS:

$$\mathbf{X}(t) = \mathbf{X}(u(t), v(t)) = (\cos u(t) \sin v(t), \sin u(t) \sin v(t), cos v(t)) +$$
$$\varepsilon[\bar{u}(t)(-\sin u(t) \sin v(t), \cos u(t) \sin v(t), 0) +$$
$$\bar{v}(t).(\cos u(t) \cos v(t), \sin u(t) \cos v(t), -sin\ v(t))]. \tag{16}$$

Then one-parameter closed motion $\mathbf{X}(t)$ correspond to a closed ruled surface in $\mathbb{R}^3$.

The most advantageous side of this method, we will determine the control points for the design is the absence of any constraint, except for the condition of being closed. So, we can achieve any closed ruled surface via this technic controlled by the user.

**Corollary 2**

$\mathbf{X}(t)$ represent a ruled surface in 3-dimensional real space and its shape changes with respect to its control points.

**4 On the Invariants of Ruled Surfaces**

In this section, properties of ruled surfaces are expressed via coordinate functions. Let take $\bar{u}, \bar{v}: \mathbb{R}^2 \mapsto \mathbb{R}, \bar{u} = \bar{u}(u,v), \bar{v} = \bar{v}(u,v), u = u(t), v = v(t)$ in the equation (13), then $\mathbf{X}(t)$ represents the one-parameter motion on DUS:

$$\mathbf{X}(t) = (\cos u(t) \sin v(t), \sin u(t) \sin v(t), \cos v(t))$$

$$+ \varepsilon \begin{bmatrix} \bar{u}(t)(-\sin u(t) \sin v(t), \cos u(t) \sin v(t), 0) \\ +\bar{v}(t).(\cos u(t) \cos v(t), \sin u(t) \cos v(t), -\sin v(t)) \end{bmatrix} \quad (17)$$

Assume that $\mathbf{X}(t) = \mathbf{X}_1(t)$, then we have

$$\mathbf{X}_1(\hat{u}(t), \hat{v}(t)) = \mathbf{X}_1(t) + \varepsilon \bar{\mathbf{X}}_1(t).$$

So equation (5) become

$$\begin{bmatrix} \bar{\mathbf{X}}'_1 \\ \bar{\mathbf{X}}'_2 \\ \bar{\mathbf{X}}'_3 \end{bmatrix} = \begin{bmatrix} -\kappa \alpha_3 & \bar{\kappa} & \kappa \alpha_1 \\ \tau \alpha_2 - \bar{\kappa} & -\kappa \alpha_3 - \tau \alpha_1 & \bar{\tau} + \kappa \alpha_2 \\ \tau \alpha_3 & -\bar{\tau} & -\tau \alpha_1 \end{bmatrix} \begin{bmatrix} \mathbf{X}_1 \\ \mathbf{X}_2 \\ \mathbf{X}_3 \end{bmatrix}.$$

Here $\alpha_1, \alpha_2$ and $\alpha_3$ can be found by following equations;

$$\bar{\mathbf{X}}_1 = \alpha_3 \mathbf{X}_2 - \alpha_2 \mathbf{X}_3, \quad \bar{\mathbf{X}}_2 = \alpha_1 \mathbf{X}_3 - \alpha_3 \mathbf{X}_1, \quad \bar{\mathbf{X}}_3 = \alpha_2 \mathbf{X}_1 - \alpha_1 \mathbf{X}_2.$$

**Corollary 3**

From the equation (17), the expressions of the dual curvature function and dual torsion function can be written by coordinate functions:

$$\kappa = \sqrt{u'^2 \sin^2 v + v'^2}$$

$$\bar{\kappa} = \frac{u'^2 \sin v (\bar{v} \cos v + \bar{u}_u \sin v) + u'v'(\bar{v}_u + \bar{u}_v \sin^2 v) + v'^2 \bar{v}_v}{\sqrt{u'^2 \sin^2 v + v'^2}}$$

$$\tau = \frac{\cos v (u'^3 \sin^2 v + 2u'v'^2) + \sin v (u'v'' + u''v')}{u'^2 \sin^2 v + v'^2}$$

$$\bar{\tau} = \sin v (\bar{u}v' - \bar{v}u') \quad (18)$$

where the subscripts indicate the partial derivative of the coordinate functions.

The distribution parameter function of the $\mathbf{X}_1(t)$ ruled surface can be found by coordinate functions

$$\delta = \frac{u'^2 \sin v(\bar{v}\cos v + \bar{u}_u \sin v) + u'v'(\bar{v}_u + \bar{u}_v \sin^2 v) + v'^2 \bar{v}_v}{u'^2 \sin^2 v + v'^2} \tag{19}$$

The directrix curve of the $\mathbf{X}_1(t)$ ruled surface can be found as follow

$$\mathbf{a}(t) = (-\bar{u}\cos u \sin v \cos v - \bar{v}\sin u, -\bar{u}\sin u \sin v \cos v + \bar{v}\cos u, \bar{u}\sin^2 v).$$

Then one can write

$$\mathbf{R}(t,w) = \mathbf{a}(t) + w\mathbf{X}_1(t), w \in \mathbb{R} \tag{20}$$

or explicitly

$$\mathbf{R}(t,w) = (-\bar{u}\cos u \sin v \cos v - \bar{v}\sin u, -\bar{u}\sin u \sin v \cos v + \bar{v}\cos u, \bar{u}\sin^2 v)$$

$$+ w(\cos u \sin v, \sin u \sin v, \cos v).$$

From equations (7) and (8) we get

$$\frac{d\mathbf{m}}{dt} = \bar{\tau}\mathbf{X}_1 + \bar{\kappa}\mathbf{X}_3 // \frac{d\mathbf{m}}{ds} = \cos\sigma(s)\mathbf{X}_1(s) + \sin\sigma(s)\mathbf{X}_3(s). \tag{21}$$

**Lemma 1**

From equation (21), we get a relation between the dual sections of dual curvature and dual torsion and the striction

$$\cot\sigma = \frac{\bar{\tau}}{\bar{\kappa}} = \frac{\sin v(\bar{u}v' - \bar{v}u')\sqrt{u'^2 \sin^2 v + v'^2}}{u'^2 \sin v(\bar{v}\cos v + \bar{u}_u \sin v) + u'v'(\bar{v}_u + \bar{u}_v \sin^2 v) + v'^2 \bar{v}_v} \tag{22}$$

**Example 1**

In example 3.1.1. let $u(t) = v(t) = t$ be. Then we have $\bar{u}(t) = 0, \bar{v}(t) = 2t$. Then from the equation (18), we can calculate the invariants of the ruled surface that correspond to these values:

$$\kappa = \sqrt{\sin^2 t + 1}$$

$$\bar{\kappa} = \frac{t\sin 2t + 3}{\sqrt{\sin^2 t + 1}}$$

$$\tau = \frac{\cos t(\sin^2 t + 2)}{\sin^2 t + 1}$$

$$\bar{\tau} = t\sin t.$$

The striction and distribution parameter of the ruled surface can be found by equations (22) and (19), respectively:

$$\cot\sigma = \frac{-t\sqrt{\sin^2 t + 1}}{t\cos t + 2\sin t}$$

$$\delta = \frac{\sin t(2t\cos t + \sin t) + \cos^2 t + 1}{\sin^2 t + 1}.$$

**4.1. Calculation of the Integral Invariants of a Closed Ruled Surface**

From equation (18) we can reformulate the integral invariants of $\mathbf{X}_1$-closed ruled surface in terms of coordinate functions:

$$l_{\mathbf{X}_1} = \oint \sin v(\bar{u}v' - \bar{v}u')\mathrm{d}t \qquad (24)$$

$$\lambda_{\mathbf{X}_1} = -\oint \frac{\cos v(u'^3\sin^2 v + 2u'v'^2) + \sin v(u'v'' + u''v')}{u'^2\sin^2 v + v'^2}\mathrm{d}t \qquad (25)$$

**4.2. Design of a Ruled Surface via Dual Spherical Bézier-Like Curves**

To indicate that the method, in section 3.2, is appropriate for calculation, we can apply it to the computer applications. From equations (15) we have the following equations:

$$u'(t) = n\sum_{i=0}^{n-1}\binom{n-1}{i}(1-t)^{n-i-1}t^i\left(\mathbf{p}_{(i+1)_1} - \mathbf{p}_{i_1}\right),$$

$$v'(t) = n \sum_{i=0}^{n-1} \binom{n-1}{i} (1-t)^{n-i-1} t^i \left( \mathbf{p}_{(i+1)_2} - \mathbf{p}_{i_2} \right),$$

$$u''(t) = n(n-1) \sum_{i=0}^{n-2} \binom{n-2}{i} (1-t)^{n-i-2} t^i \left( \mathbf{p}_{(i+2)_1} - 2\mathbf{p}_{(i+1)_1} + \mathbf{p}_{i_1} \right),$$

$$v''(t) = n(n-1) \sum_{i=0}^{n-2} \binom{n-2}{i} (1-t)^{n-i-2} t^i \left( \mathbf{p}_{(i+2)_2} - 2\mathbf{p}_{(i+1)_2} + \mathbf{p}_{i_2} \right).$$

Thus, if we put these values into former equations of the ruled surface then they belong to the control points. We can calculate the these values for some points specifically since these calculations do not fit in this paper. For example, let the parameter value be zero, i.e. $t = 0$, then we have following values:

$$u'(0) = n(\mathbf{p}_{1_1} - \mathbf{p}_{0_1}), \quad v'(0) = n(\mathbf{p}_{1_2} - \mathbf{p}_{0_2}),$$

$$u''(0) = n(n-1)(\mathbf{p}_{2_1} - 2\mathbf{p}_{1_1} + \mathbf{p}_{0_1}),$$

$$v''(0) = n(n-1)(\mathbf{p}_{2_2} - 2\mathbf{p}_{1_2} + \mathbf{p}_{0_2}).$$

So we can calculate the curvature, torsion, distribution parameter and the striction at this point.

$$\kappa(0) = n \sqrt{(\mathbf{p}_{1_1} - \mathbf{p}_{0_1})^2 \sin^2(\mathbf{p}_{0_2}) + (\mathbf{p}_{1_2} - \mathbf{p}_{0_2})^2}$$

$$\bar{\kappa}(0) = \frac{n \left[ (\mathbf{p}_{1_1} - \mathbf{p}_{0_1})^2 \sin(\mathbf{p}_{0_2}) \left( \bar{v} \cos(\mathbf{p}_{0_2}) + \bar{u}_u \sin(\mathbf{p}_{0_2}) \right) \right]}{\sqrt{(\mathbf{p}_{1_1} - \mathbf{p}_{0_1})^2 \sin^2(\mathbf{p}_{0_2}) + (\mathbf{p}_{1_2} - \mathbf{p}_{0_2})^2}}$$

$$+ \frac{n \left[ (\mathbf{p}_{1_1} - \mathbf{p}_{0_1})(\mathbf{p}_{1_2} - \mathbf{p}_{0_2}) \left( \bar{v}_u + \bar{u}_v \sin^2(\mathbf{p}_{0_2}) \right) + (\mathbf{p}_{1_2} - \mathbf{p}_{0_2})^2 \bar{v}_v \right]}{\sqrt{(\mathbf{p}_{1_1} - \mathbf{p}_{0_1})^2 \sin^2(\mathbf{p}_{0_2}) + (\mathbf{p}_{1_2} - \mathbf{p}_{0_2})^2}}$$

$$\tau(0) = \frac{n\left[\cos(\mathbf{p}_{0_2})(\mathbf{p}_{1_1} - \mathbf{p}_{0_1})\left((\mathbf{p}_{1_1} - \mathbf{p}_{0_1})^2 \sin^2(\mathbf{p}_{0_2}) + 2(\mathbf{p}_{1_2} - \mathbf{p}_{0_2})^2\right)\right]}{(\mathbf{p}_{1_1} - \mathbf{p}_{0_1})^2 \sin^2(\mathbf{p}_{0_2}) + (\mathbf{p}_{1_2} - \mathbf{p}_{0_2})^2}$$

$$+ \frac{(n-1)\sin(\mathbf{p}_{0_2})\left((\mathbf{p}_{1_1} - \mathbf{p}_{0_1})(\mathbf{p}_{2_2} - 2\mathbf{p}_{1_2} + \mathbf{p}_{0_2}) + (\mathbf{p}_{1_2} - \mathbf{p}_{0_2})(\mathbf{p}_{2_1} - 2\mathbf{p}_{1_1} + \mathbf{p}_{0_1})\right)}{(\mathbf{p}_{1_1} - \mathbf{p}_{0_1})^2 \sin^2(\mathbf{p}_{0_2}) + (\mathbf{p}_{1_2} - \mathbf{p}_{0_2})^2}$$

$$\bar{\tau}(0) = n\sin(\mathbf{p}_{0_2})\left(\bar{u}(\mathbf{p}_{1_2} - \mathbf{p}_{0_2}) - \bar{v}(\mathbf{p}_{1_1} - \mathbf{p}_{0_1})\right)$$

$$\delta(0) = \frac{n\left[(\mathbf{p}_{1_1} - \mathbf{p}_{0_1})^2 \sin(\mathbf{p}_{0_2})\left(\bar{v}\cos(\mathbf{p}_{0_2}) + \bar{u}_u\sin(\mathbf{p}_{0_2})\right)\right]}{(\mathbf{p}_{1_1} - \mathbf{p}_{0_1})^2 \sin^2(\mathbf{p}_{0_2}) + (\mathbf{p}_{1_2} - \mathbf{p}_{0_2})^2}$$

$$\cot(\sigma(0)) = \frac{\left(\bar{u}(\mathbf{p}_{1_2} - \mathbf{p}_{0_2}) - \bar{v}(\mathbf{p}_{1_1} - \mathbf{p}_{0_1})\right)\sqrt{(\mathbf{p}_{1_1} - \mathbf{p}_{0_1})^2 \sin^2(\mathbf{p}_{0_2}) + (\mathbf{p}_{1_2} - \mathbf{p}_{0_2})^2}}{(\mathbf{p}_{1_1} - \mathbf{p}_{0_1})^2\left(\bar{v}\cos(\mathbf{p}_{0_2}) + \bar{u}_u\sin(\mathbf{p}_{0_2})\right)}$$

For *t* = 1, it can be calculated in the same way.

We can show the design of a ruled surface in an example:

**Example 2**

Let the domain of RUS be $\mathcal{B} = \{(u,v): 0 \leq u \leq \pi, \; 0 \leq v \leq 2\pi\}$ and choose the control points on this domain as follows:

p(0) = $[\pi/8, \pi/4]$, p(1) = $[\pi/8, 3\pi/8]$, p(2) = $[3\pi/8, 3\pi/8]$, p(3) = $[3\pi/8, \pi/4]$,

p(4) = $[3\pi/8, \pi/8]$, p(5) = $[\pi/8, \pi/8]$, p(6) = p(0).

Then the Bézier curve on this domain can be expressed as:

$$\mathbf{b}(t) = \sum_{i=0}^{6} \binom{6}{i}(1-t)^{n-i} t^i \mathbf{p}_i$$

(Fig 4.3).

So we can map this curve to the RUS by

$$\mathbf{X}(t) = \left(\cos u(t) \sin v(t), \sin u(t) \sin v(t), \cos v(t)\right)$$

parametrization (Fig 4.4).

In addition to describing the dual closed Bézier-like curve we can take $\bar{u} = u - v$ ve $\bar{v} = u + v$ and if we put the control points in equation (15) we get a dual closed Bézier-like curve on DUS. Corresponding ruled surface is shown in figure (Fig 4.5). Furthermore, integral invariants of this ruled surface can be calculated as,

$\lambda_{\mathbf{X}(t)} = 1.419793061$

$l_{\mathbf{X}(t)} = 0.3381414433.$

**5 Conclusion and Future Work**

Several scientists investigated spherical Bézier curves by using deCasteljau-type algorithms. Those algorithms run a large number of inputs since they have to count the control points and also the angles among position vectors of the control points. This situation causes challenges in view of computation. Technically, several mathematical software can not compute the spherical Bézier curve. These methods are not suitable for us because one of our main purposes is to calculate the integral invariants of the closed ruled surface. We have developed a new method that is more efficient computable than the others in order to construct spherical Bézier-like curves.

In this paper, we have designed spherical Bézier-like curves via mapping the planar Bézier curves to the surface of RUS by unit sphere mapping. Naturally, coordinate functions emerge since we obtain ruled surfaces from a synectic congruence. Therefore, we obtained devisable ruled surfaces with respect to the control points. The control points can be chosen either on the plane(domain) or on the surface of RUS since the unit sphere mapping is one-to-one. The user can design the ruled surfaces with this method. Besides, the method is more suitable and faster for computation than the former methods.

Obviously, we can construct the dual spherical B-spline curves via constructing B-spline curves in the domain and mapping to the RUS. Thus, the design of the corresponding ruled surface belongs to the control points and knots of the spline curve.

Plane kinematics can be studied by this method and therefore results of spherical kinematics can be viewed. For example, Holditch-type theorems(plane and spherical) can be investigated in terms of control points.

In addition to Euclidean space, the ruled surfaces in hyperbolic space and Lorentz space can be designed by control points. So, timelike and spacelike ruled surfaces can be investigated by control points.

**REFERENCES**

[1] E. Study, 1903, *Geometrie der Dynamen*, Leipzig.

[2] Hoschek, J., 1973, Integralinvarianten von Regelflachen. *Arch. Math*. *XXIV*, pp. 218-224.

[3] Gürsoy, O., 1990, The dual angle of pitch of a closed ruled surface, *Mechanism and Machine Theory*, *Great Britain*, Vol.25, No:2, pp. 131-140, doi:10.1016/0094-114X(90)90114-Y.

[4] Shoemake K., 1985. Animating rotation with quaternion curves. *Proceedings of SIGGRAPH'85*, July 22-26 San Francisco, CA, 19 (3), 245-254.

[5] Crouch, P., Kun, G., Silva Leite, F., 1999, The de Casteljau algorithm on Lie groups and spheres, *Journal of Dynamical and Control Systems*, Vol. 5, pp. 397-429, 10.1023/A:1021770717822.

[6] Ge, Q.J., Ravani, B., 1994, Geometric construction of Bézier motions, *Journal of Mechanical Design*, Vol. 116, pp. 749-755, doi:10.1115/1.2919446.

[7] Ge, Q.J., Ravani, B., 1994, Computer aided geometric design of motion interpolants, *Journal of Mechanical Design*, Vol. 116, pp. 756-762, doi:10.1115/1.2919447.

[8] Yayun, Z., Schulze, J., Schaffler, S., 2012, Dual Spherical Spline: a New Representation of Ruled Surface Optimization, *XII. International Conference on Control, Automation, Robotics&Vision*, China, 1193-1198.

[9] Taş F., "A METHOD OF DETERMINING DUAL UNIT SPHERICAL BÉZIER CURVES", Journal of Logic, Mathematics and Linguistics in Applied Sciences, pp.1-6, 2016

[10] W. K. Clifford, 1873, *Preliminary sketch of bi-quaternions*, Proc. London Math. Soc. Vol. 4, pp. 381-395.

[11] Blaschke, W., 1949, *Diferansiyel geometri dersleri*, İstanbul Üniversitesi Yayınları, No. 433, İstanbul.

[12] Pottmann, H., Wallner, J., 2001, *Computational line geometry*, Mathematics and Visualization, Springer-Verlag Berlin Heidelberg, e-ISBN 978-3-642-04018-4.

[13] Guggenheimer, H.W., 1977, *Differential geometry*, Dover Publications, Inc., New York.
21

# Figure Captions List

Fig. 1          Dual representation of a line

Fig. 2          Closed Bézier curve on the domain

Fig. 3          Closed spherical Bézier curve on RUS

Fig. 4          The closed ruled surface with its striction curve

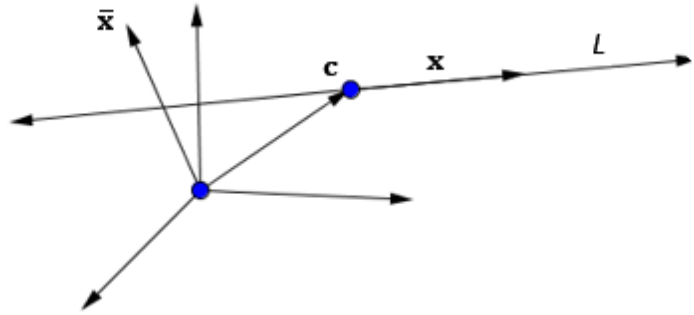

Fig. 1 Dual representation of a line

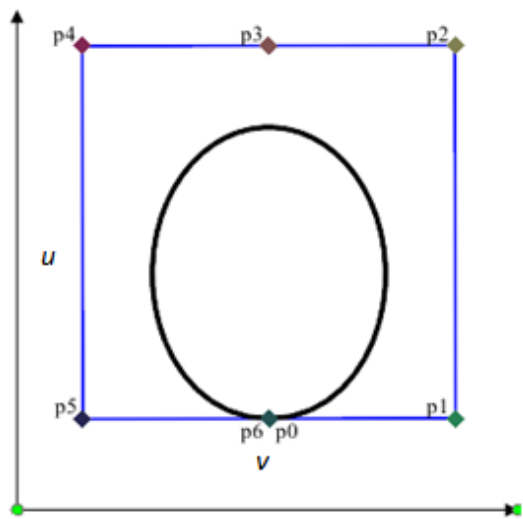

Fig. 2 Closed Bézier curve on the domain

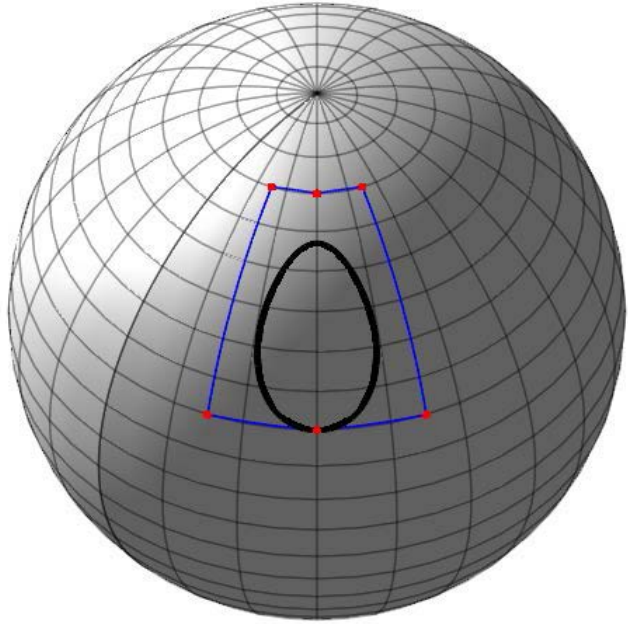

Fig. 3 Closed spherical Bézier curve on RUS

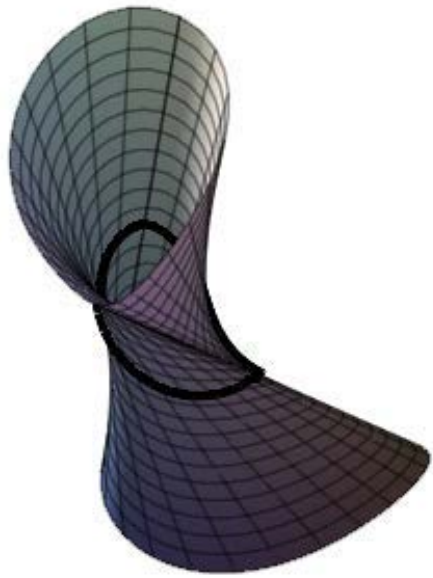

Fig. 4 The closed ruled surface with its striction curve